\documentstyle[prl,twocolumn,aps,epsfig]{revtex}
\begin{document}
\title{Statistics of Wave Functions in Coupled Chaotic Systems}
\author{A.~Tschersich$^{1}$ and K.B.~Efetov$^{1,2}$}
\address{$^{1}$Fakult\"at f\"ur Physik und Astronomie, Ruhr-Universit\"at-Bochum,\\
Universit\"atsstr.~150, 44780 Bochum, Germany\\
$^{2}$L.D.~Landau Institute for Theoretical Physics, Moscow, Russia}
\date{\today{}}
\maketitle
\draft

\begin{abstract}
Using the supersymmetry technique, we calculate the joint distribution of local densities of electron wavefunctions in two coupled disordered or chaotic quantum billiards. We find novel spatial correlations that are absent in a single chaotic system. Our exact result can be interpreted for small coupling in terms of the hybridization of eigenstates of the isolated billiards. We show that the presented picture is universal, independent of microscopic details of the coupling.
\end{abstract}


\bigskip

% 05.45.Mt Semiclassical chaos (``quantum chaos'')
% 73.23.-b Mesoscopic systems
% 73.20.Dx Electron states in low-dimensional structures
% 73.23.Hk Coulomb blockade; single- electron tunneling

In disordered or chaotic systems the dynamics is in general too complex to find exact eigenstates for a given disorder or boundary configuration. Therefore, one is interested in statistical properties of physical quantities like wave functions, energy levels, etc. The complete information about these quantities can be extracted from distribution functions that can be calculated either numerically or analytically. Various methods of studying the statistics have been developed recently and this gave the possibility to achieve a good understanding of many disordered and ballistic chaotic systems (for a recent review see, e.g.\ \cite{EfetovBook,Falko,Beenakker0,Guhr,Mirlin99}).

Now it is well established that under very general conditions the statistics of chaotic systems is universal and well described by random matrix theory (RMT) or the zero-dimensional version of the supermatrix $\sigma $-model. For the distribution function of the amplitudes of wave functions one obtains within this approximation the famous Porter-Thomas distribution \cite{PorterThomas}, which is simply a Gaussian distribution of the amplitudes. Then, fluctuations of the wave functions at different points are statistically independent.

In this Letter, we present results from studying statistical properties of wave functions of two coupled chaotic systems. To be more specific, we consider two quantum chaotic billiards coupled in such a way that electrons (or electromagnetic waves) can penetrate from one billiard into the other. Studying such types of models is definitely interesting, because they can be relevant for double quantum dot systems recently fabricated on the basis of a 2D electron gas and investigated experimentally \cite{WaughEtAl}. One can also imagine that in the nearest future microwave experiments on double billiards will be performed. However, the importance of studying coupled chaotic systems is much more general and results obtained can be relevant e.g.\ for systems of weakly coupled complex atoms and molecules. To the best of our knowledge, neither numerical nor analytical results from a study of wave functions of coupled chaotic systems exist in the literature.

Wave functions are the most interesting objects in coupled systems, because even at very small coupling they can drastically change if any two levels of the corresponding isolated systems are close to each other. Statistics of such hybridized wave functions can be quite unusual, but, at the same time, possess interesting universal features.

The probability for the local density $V|\phi _{\alpha }(x)|^{2}$ of a state with wave function $\phi _{\alpha }(x)$ and energy $\epsilon_{\alpha }$ as well as its spatial correlations can be described by the joint distribution function (JDF) $f_{2}(p_{1},p_{2})$, defined as: 
\begin{equation}
  f_{2}(p_{1},p_{2})=\left\langle \Delta \sum_{\alpha }\delta \left( \epsilon -\epsilon _{\alpha }\right) \prod_{i=1,2}\delta \left( p_{i}-V|\phi _{\alpha}(x_{i})|^{2}\right) \right\rangle,  \label{eqn_JDF}
\end{equation}
where the brackets $\langle \dots \rangle $ denote disorder averaging, $\Delta =\left( \nu V\right) ^{-1}$ is the mean level spacing, $V$ is the volume of the total system and $\nu $ is the average density of states. The function $f_{2}(p_{1},p_{2})$ is the probability that the local densities at energy $\epsilon $ are equal to $p_{1}$ and $p_{2}$ at points $x_{1}$ and $x_{2}$ correspondingly.

If $p_{1}$ and $p_{2}$ are not very large the statistics of the wave functions is universal and does not depend on whether the chaotic motion is due to disorder or a non-integrable shape of the billiard. The direct calculation using the supersymmetry technique or RMT gives for single billiards in the unitary case:
\begin{equation}
  f_{2}(p_{1},p_{2})=\exp (-p_{1})\,\exp (-p_{2}),  \label{a1}
\end{equation}
which is just the Porter-Thomas distribution\cite{PorterThomas}. Proper formulae can be written for the other ensembles but we concentrate here on studying a chaotic system with broken time reversal symmetry. This case is most simple for calculations but the interesting effects obtained below are general. The unitary ensemble is easily realized in quantum dots by applying a magnetic field. As concerns microwave cavities, time reversal invariance can be broken by a special preparation of the cavity walls \cite{Stoeckmann}.

The function $f_{2}\left( p_{1},p_{2}\right) $, Eq.~(\ref{eqn_JDF}), characterizes correlations of the amplitudes of the electromagnetic waves in microwave experiments and can be measured directly \cite{KudrolliKidambiSridhar}. At the same time, this function determines the statistics of the conductance peak heights $G_{\max }$ in quantum dots in the regime of Coulomb blockade. The conductance $G_{\max }$ can be expressed through the tunnel rates of the contacts to external leads, $\Gamma _{i}\propto \left| \phi \left( x_{i}\right) \right| ^{2}$ \cite{Beenakker91,JalabertStoneAlhassid}: 
\begin{equation}
  G_{{\rm max}}\propto \frac{e^{2}}{T}\frac{\Gamma _{1}\Gamma _{2}}{\Gamma_{1}+\Gamma _{2}}.  \label{a2}
\end{equation}
Eq.~(\ref{a2}) is valid for $\Gamma _{i}\ll T\ll \Delta $, where $T$ is temperature and $\Delta $ is the mean level spacing in a single dot. Then we have for the distribution function $P\left( g\right) $ of the dimensionless conductance $g=T/(\Gamma e^2)\, G_{\max }$:
\begin{equation}
  P(g)=\int {\rm d}p_{1}\,{\rm d}p_{2}\,\delta \left( g-\frac{p_{1}p_{2}}{p_{1}+p_{2}}\right) \,f_{2}(p_{1},p_{2}).  \label{a3}
\end{equation}
The relevance of Eq.~(\ref{a1}) for conductance fluctuations in single dots has been confirmed in experiments \cite{Marcus}. Unfortunately, conductance fluctuations were not studied in an experiment on double quantum dots \cite{WaughEtAl}. In a recent theoretical work \cite{KaminskiGlazman} the averaged resonance conductance and its fluctuations have been considered for two coupled dots in the limit of weak coupling based on the assumption that wavefunctions of the corresponding isolated dots are Porter-Thomas distributed even near the point-contact. In contrast, we present a general microscopic derivation of the function $f_{2}\left(p_{1},p_{2}\right)$, Eq.~(\ref{eqn_JDF}), for two coupled quantum chaotic billiards for arbitrary coupling. We find novel universal correlations that do not exist in a single billiard.

\begin{figure}[h]
  \begin{center}
    \epsfig{file=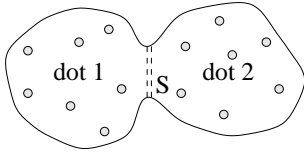,height=2cm}
    \caption{\label{fig_dots} Two quantum billiards are coupled along some extended tunneling surface $S$.}
  \end{center}
\end{figure}
The system we consider is represented in Fig.~(\ref{fig_dots}). The Hamiltonian $H$ can be written as:
\begin{equation}
  H=H_{1}+H_{2}+H_{T},  \label{a4}
\end{equation}
where $H_{1}$ and $H_{2}$ correspond to the isolated dots:
\begin{equation}
  H_{i}=\sum_{\alpha ;i=1,2}\epsilon _{\alpha }^{i}c_{\alpha }^{i+}\, c_{\alpha}^{i}.  \label{a5}
\end{equation}
In Eq.~(\ref{a5}), $\epsilon _{\alpha }^{i}$ are the single particle energies of dot $i$, and $c^{+}$ and $c$ are creation and annihilation operators. We want to emphasize that the billiards are not identical and can have arbitrary shapes and configurations of disorder. The coupling between the billiards is described by a tunneling Hamiltonian $H_{T}$ that can be written as:
\begin{equation}
  H_{T}=\sum_{\alpha ,\beta }t_{\alpha \beta }c{_{\alpha }^{1}}^{\!+} c_{\beta}^{2}+{\rm c.c.}  \label{a6}
\end{equation}
Assuming that the particles tunnel through a surface $S$ between the two dots, the tunneling matrix element $t_{\alpha \beta }$ takes the form:
\begin{equation}
  t_{\alpha \beta }=t_{0}\int_{S}{\rm d}^{d-1}x\,\sqrt{V_{1}V_{2}} {\phi_{\alpha }^{1}}^{\!\ast }(x)\,\phi _{\beta }^{2}(x).  \label{a7}
\end{equation}
In Eq.~(\ref{a7}) we integrate over the surface $S$, $V_{i}$ is the volume of dot $i$, $t_{0}$ is the tunneling rate per unit surface and $\phi_{\alpha }^{i}(x)$ is the single particle wave function of state $\alpha$ at a point $x$ of the tunnel contact in dot $i$.

Starting with the Hamiltonian, Eqs.~(\ref{a4}-\ref{a7}), we perform calculations using the supersymmetry technique \cite{EfetovBook}. Following this method the JDF can be written in a form given by an integral over supermatrices $Q_{1}$, $Q_{2}$ \cite{EfetovFalko96}: 
\begin{eqnarray}
  f_{2}(p_{1},p_{2}) &=&\lim_{{\delta {\rightarrow 1} \atop {\gamma \rightarrow 0}}} \frac{{\rm d}}{{\rm d}\delta }\frac{\delta }{4} \int_{0}^{1}\!\!\! {\rm d}t\, {\rm d}Q_{1} {\rm d}Q_{2}\, \delta \left(p_{1}+\frac{\gamma t\delta }{2\Delta }{\rm Str}(\Pi_{ab}Q_{1})\right)  \nonumber \\
  &&\hspace{-1.5cm}\delta \left(p_{2}-\frac{\gamma (1-t)\delta }{2\Delta } {\rm Str}(\Pi _{rb}Q_{2})\right)\, {\cal C}[Q_{i}] \exp \left( -F\left[ Q\right] \right).  
\label{a8}
\end{eqnarray}
The free energy $F=F_{0}+F_{T}$ contains the free part $F_{0}$ for the single dots:
\begin{equation}
  F_{0}=-\frac{\gamma }{4}\sum_{i=1,2}\Delta _{i}^{-1}{\rm Str}(\Lambda Q_{i}),
\label{a9}
\end{equation}
and the term $F_{T}$ describing the coupling between them:
\begin{equation}
  F_{T}=-\frac{\alpha }{4}{\rm Str}(Q_{1}Q_{2})  \label{a10}
\end{equation}
The function ${\cal C}[Q_{i}]$ in Eq.~(\ref{a8}) is:    
\[
  {\cal C}[Q_{i}]={\rm Str}\left( \left( \Pi _{af}-\Pi _{rf}\right) \left(D_{1}Q_{1}+D_{2}Q_{2}\right) \right) ,
\]
$\Delta _{i}=\left( \nu V_{i}\right) ^{-1}$ is the mean level spacing in dot $i$, $D_{i}=\Delta /\Delta _{i}$ and $\Delta ^{-1}=\nu \left(V_{1}+V_{2}\right) $. The projectors $\Pi _{ij}$ select certain elements of the supermatrices $Q_{i}$, $\Lambda ={\rm diag}\{1,1,1,1,-1,-1,-1,-1\}$, ${\rm Str}$ is the supertrace (for more details of the notations see \cite{EfetovBook}).

The parameter $\alpha \propto \lambda _{F}^{d-1}S|t_{0}|^{2}\left( \Delta_{1}\Delta _{2}\right) ^{-1}$ is the dimensionless coupling constant, with $\lambda _{F}$ being the Fermi wavelength. In $F_{T}$ we have kept only the term of lowest order in $\alpha $. Other terms of order $\alpha ^{n}$ are smaller by a factor $\sim \left( \lambda _{F}^{d-1}/S\right) ^{n-1}$ and this corresponds to many-channel tunneling between the dots. Therefore, as long as $\lambda _{F}^{d-1}/S\ll 1$, we can consider not only small couplings but also $\alpha \geq 1$. On the other hand, neglecting higher order terms for point contacts is correct only for $\alpha \ll 1$. Note, that in the limit $\alpha \rightarrow \infty $ fluctuations of $\left( Q_{1}-Q_{2}\right) $ are suppressed and one comes to the model for a single dot with volume $V=V_{1}+V_{2}$. The opposite limit, $\alpha \rightarrow 0$, can only be performed after taking $\gamma \rightarrow 0$.

To perform the integration we use the standard parametrization \cite{EfetovBook}. All manipulations are not very difficult for the unitary ensemble. After a rotation of $Q_{2}$ only one set of Grassmann variables remains in the exponent, which is convenient for integrating out the Grassmann variables. Due to the presence of the delta functions, the integration over the non-compact bosonic sector is simplified in the limit $\gamma \rightarrow 0$ and we get the final result valid for arbitrary values of the coupling $\alpha $:
\begin{eqnarray}
  \lefteqn{f_{2}(p_{1},p_{2})=\sqrt{\alpha }\,\exp \left(-D_{1}p_{1}-D_{2}p_{2}-\beta \right) }  \label{eqn_exactresult} \\
  && \sum_{i\neq k}D_{i}\left( q_{k}+\alpha \right) ^{-1/2}\left[A_{i}+B_{i}\left( q_{i}+\alpha \right) +C_{i}\left( q_{i}+\alpha \right) ^{2}\right] .\nonumber
\end{eqnarray}
In all the expressions $i\neq k$ and both indices can take values $1$ and $2$. We have defined for compact notation $q_{i}=2D_{k}p_{i}$, $\beta =\sqrt{(q_{1}+\alpha )(q_{2}+\alpha )}$ and:
\begin{eqnarray*}
  A_{i} &=&3D_{1}D_{2}m+\left( 3D_{k}^{2}+D_{i}^{2}\right) n, \\
  B_{i} &=&\left( D_{i}^{2}m+3D_{1}D_{2}n\right)\left( 1+\beta ^{-1}\right)\beta ^{-1}, \\
  C_{i} &=& D_{i}^{2}n\left( 1+3\beta ^{-1}+3\beta ^{-2}\right) \beta ^{-2}, \\
  m &=&\cosh \alpha -\frac{\sinh \alpha }{2\alpha },\quad n=\frac{1}{2}\sinh\alpha .
\end{eqnarray*}

Eq.~(\ref{eqn_exactresult}), although being exact, is rather complicated and we discuss now its asymptotics. In the limit $\alpha \rightarrow 0$ corresponding to almost isolated quantum dots the JDF vanishes for all $p_{1}$, $p_{2}$ excluding the case when either $p_{1}$ or $p_{2}$ is zero, where it diverges. The divergencies are present in the terms proportional to $\beta ^{-4}$ in $C_{i}$ and the terms proportional to $\beta ^{-2}$ in $B_{i}$. Then we find:
\begin{equation}
  f_{2}(p_{1},p_{2})=D_{2}^{2}\,\delta(p_{1})\,e^{-D_{2}p_{2}}+D_{1}^{2}\,\delta (p_{2})\,e^{-D_{1}p_{1}}.
\label{a11}
\end{equation}
The particle is localized either in dot $1$, then $p_{2}=0$ and $p_{1}$ is distributed via Porter-Thomas, or vice versa. The factors $D_{i}$ appear, because the dimensionless densities $p_{i}$ have been normalized with respect to the volume of the total system and not of the corresponding single dot.

In the opposite limit $\alpha \rightarrow \infty $ of strong coupling between the dots we can easily obtain from Eq.~(\ref{eqn_exactresult}) the Porter-Thomas distribution, Eq.~(\ref{a1}), for the total system with volume $V$.

So, in the limit of strongly coupled and isolated billiards we obtain the Porter-Thomas distribution. In both cases fluctuations of the amplitudes of the wave functions at different points are Gaussian and not correlated. But what happens if the coupling constant $\alpha $ is finite although small? Can one have for such $\alpha $ correlations between the wave functions at different points within one dot or, maybe, correlations between different dots?

The answer to this question can be found taking in Eq.~(\ref{eqn_exactresult}) the limit $\alpha \ll 1$ and $p_{1},p_{2}\gg \alpha $, which gives: 
\begin{eqnarray}
  f_{2}(p_{1},p_{2}) &\approx &\sqrt{\frac{\alpha }{8}}\left(D_{1}D_{2}\right) ^{3/2}\exp \left[ -\left( \sqrt{u_{1}}+\sqrt{u_{2}}\right)^{2}\right]   \label{eqn_JDFtwolevelapprox} \\
  &&\hspace{-1.8cm}\left[ 3\left( u_{1}^{-1/2}+u_{2}^{-1/2}\right) +\left(\frac{1}{2}+\sqrt{u_{1}u_{2}}\right) \left( u_{1}^{-3/2}+u_{2}^{-3/2}\right) \right],   \nonumber
\end{eqnarray}
where $u_{i}=D_{i}p_{i}$. We see, that the exponential cannot be factorized as in the case of the ordinary Porter-Thomas distribution, which demonstrates novel correlations that are absent in a single dot. This quite remarkable result means that putting a weakly penetrable wall in a quantum billiard yields correlations of the wave functions across the wall, although without the wall they would be absent. Of course, making the wall less and less penetrable reduces correlations due to the prefactor $\sqrt{\alpha }$ in Eq.~(\ref{eqn_JDFtwolevelapprox}) but this decay is slow.

The form of the function $f_{2}\left( p_{1},p_{2}\right) $, Eq.~(\ref{eqn_JDFtwolevelapprox}), is surprisingly universal with only one sample specific parameter $\alpha $ describing the tunneling. It is natural to assume that such universality follows from some general physical principles. Indeed, the correlations between the wave functions of the chaotic billiards is a consequence of the quantum mechanical hybridization of states. This hybridization is most effective when the coupling between the billiards is weak and when any two levels of different billiards are close to each other.

Let us show that Eq.~(\ref{eqn_JDFtwolevelapprox}) can be obtained in the limit $\alpha \ll 1$, $p_{1,2}\gg 1$ from this picture. The derivation is not as general and we are able to reproduce Eq.~(\ref{eqn_JDFtwolevelapprox}) for point contacts only. Instead of Eq.~(\ref{a7}) we consider a tunneling point contact at $x_{t}$: 
\begin{equation}
  t_{\alpha \beta }=t_{0}\,\sqrt{V_{1}V_{2}}\,{\phi _{\alpha }^{1}}^{\!\ast}(x_{t})\,\phi _{\beta }^{2}(x_{t}).  \label{a12}
\end{equation}
The dimensionless coupling constant is now $\alpha \propto \frac{|t_{0}|^{2}}{\Delta _{1}\Delta _{2}}$. We assume that the main contribution comes from configurations where two levels of the dots $1$ and $2$ are close to each other and so we consider only two eigenstates with energy $\xi ^{1}$ in dot $1$ and energy $\xi ^{2}$ in dot $2$. Including the tunneling results in a hybridization of the states. Following \cite{KaminskiGlazman} we write the energies of the hybridized states in first order of degenerate perturbation theory: 
\begin{equation}
  \epsilon _{\pm }=\xi _{+}\pm \sqrt{\xi _{-}^{2}+|t_{12}|^{2}},  \label{a13}
\end{equation}
where $\xi _{+}=\frac{1}{2}(\xi ^{1}+\xi ^{2})$ and $\xi _{-}=\frac{1}{2}(\xi ^{1}-\xi ^{2})$. The local densities $|\phi _{\pm }(x_{1,2})|^{2}$ in dots $1,2$ at the points $x_{1}$ and $x_{2}$ near the contact can now be expressed in terms of the eigenfunctions $\phi ^{1}\left( x_{1}\right) $ and $\phi ^{2}\left( x_{2}\right) $ of the isolated dots: 
\begin{eqnarray}
  |\phi _{\pm }(x_{1})|^{2} &=&\left[ 1+{\cal A}_{\mp }\right] ^{-1}\,|\phi^{1}(x_{1})|^{2},  \label{a14} \\
  |\phi _{\pm }(x_{2})|^{2} &=&\left[ 1+{\cal A}_{\mp }^{-1}\right]^{-1}\,|\phi ^{2}(x_{2})|^{2},  \nonumber
\end{eqnarray}
where ${\cal A}_{\mp }=|t_{12}|^{-2}\left( \xi _{-}\mp \sqrt{\xi_{-}^{2}+|t_{12}|^{2}}\right) ^{2}$.

In order to calculate the function $f_{2}\left( p_{1},p_{2}\right) $, the local densities $|\phi _{\pm }(x_{1,2})|^{2}$, Eqs.~(\ref{a13}-\ref{a14}), should be substituted into Eq.~(\ref{eqn_JDF}) and the latter averaged over disorder. Instead of doing so, we replace the averaging over the disorder by an averaging over the level mismatch $\xi _{-}$ and the densities $p_{i}=V_{i}|\phi ^{i}(x_{i})|^{2}$ in all points. Assuming that the densities in the isolated dots are not correlated, being Porter-Thomas distributed, and the level mismatch distribution being given by a function $s$: 
\begin{equation}
  \Delta ^{-1}s(\xi _{-}/\Delta )\,{\rm d}\xi _{-},  \label{a15}
\end{equation}
we can perform the averaging for the function $f_{2}\left(p_{1},p_{2}\right)$. Now Eq.~(\ref{eqn_JDF}) becomes: 
\begin{eqnarray}
  f_{2}(p_{1},p_{2}) &=&\frac{1}{\Delta }\sum_{\pm }\int_{0}^{\infty }\left(\prod_{i=1}^{4}{\rm d}q_{i}\,e^{-q_{i}}\right) \int_{-\infty }^{\infty }\!\!{\rm d}\xi _{-}s\left( \xi _{-}/\Delta \right)   \nonumber \\
  &&\hspace{-1.8cm}\,\delta \left( p_{1}-D_{1}^{-1}[1+{\cal A}_{\mp}]^{-1}q_{3}\right) \,\delta \left( p_{2}-D_{2}^{-1}\left[ 1+{\cal A}_{\mp}^{-1}\right] ^{-1}q_{4}\right)   \nonumber
\end{eqnarray}
where ${\cal A}_{\mp }=\left( t_{0}^{2}q_{1}q_{2}\right) ^{-1}\left( \xi_{-}\mp \sqrt{\xi _{-}^{2}+t_{0}^{2}q_{1}q_{2}}\right) ^{2}$.

Evaluation of the integrals is straightforward. First we integrate over $q_{3},q_{4}$ and then over $\xi _{-}$. In the limit $p_{1},p_{2}\gg\alpha $, $\alpha \ll 1$ the main contribution to the integral comes from $\xi _{-}\sim t_0$ and we can replace the function $s\left( \xi _{-}/\Delta\right) $ by the value $s\left( 0\right) $. This shows that the result is independent of the distribution of the mismatch $\xi _{-}$. The remaining integrals can be performed easily and we reproduce Eq.~(\ref{eqn_JDFtwolevelapprox}) up to a constant factor of order unity which cannot be found from such a consideration.

Nevertheless, this simple evaluation cannot replace the explicit derivation of the general Eq.~(\ref{eqn_exactresult}) with the supersymmetry technique, because Eq.~(\ref{eqn_exactresult}) is valid for an arbitrary $\alpha $ and $p_{1,2}$. Even in the limit $\alpha \ll 1$, $p_{1,2}\gg 1$ the evaluation based on the two-level approximation was done for point contacts only. If we used instead of Eq.~(\ref{a12}) the more general Eq.~(\ref{a7}), we would have in the expression for $|t_{12}|^{2}$ combinations of the type $\prod_{i}{\phi }^{i\ast }(x)\,\phi ^{i}(x^{\prime })$ with $|x-x^{\prime }|\sim\lambda _{F}$. However, the Porter-Thomas distribution is not applicable for describing wave functions at different points separated by atomic distances.

In contrast, Eq.~(\ref{eqn_exactresult}) does not depend on the structure of the tunnel contact and contains as the only parameter $\alpha $. We have, therefore, identified universal statistics of hybridized levels that is independent of microscopic details and follows only from the existence of a coupling between two otherwise statistically independent systems.

\begin{figure}[tbp]
  \begin{center}
    \epsfig{file=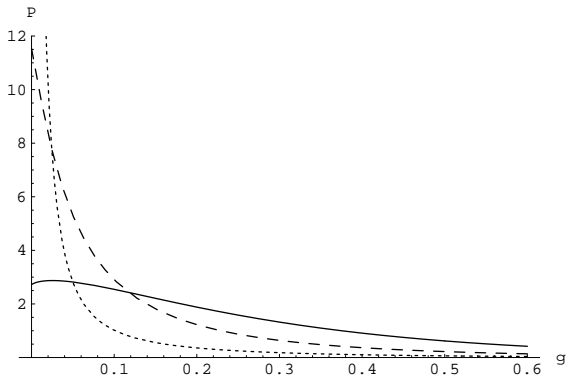,height=5cm}
    \caption{\label{fig_condist} Conductance peak distribution $P(g)$ for different couplings $\protect\alpha=0.01$ (dotted line), $0.1$ (dashed line), and $1$ (solid line).}
  \end{center}
\end{figure}
As a natural application of the general formula, Eq.~(\ref{eqn_exactresult}), let us calculate the distribution function for the peaks of the conductance in a double dot structure in the regime of Coulomb blockade. Substituting Eq.~(\ref{eqn_exactresult}) into Eq.~(\ref{a3}) we can evaluate immediately the distribution function $P\left( g\right) $, where $g$ is the dimensionless conductance. The result of the computation is represented in Fig.~(\ref{fig_condist}). It should be mentioned that Eq.~(\ref{a3}) and, hence, the plot in Fig.~(\ref{fig_condist}) are valid for a symmetric setup with both dots of the same volume and equal couplings to the leads (generalizing to an asymmetric situation can easily be done). For small couplings $\alpha \ll 1$ the statistical properties of the peak conductance are correctly described by the two level approximation \cite{KaminskiGlazman}. In this region small values of the conductance peaks are most probable. For large $\alpha $ the distribution is almost flat.

In conclusion, we derived the joint distribution function of the amplitudes of wave functions for coupled chaotic systems in the unitary ensemble. We have discovered novel universal correlations that are absent in a single chaotic system and presented simple physical arguments explaining the origin of these correlations. The results obtained can be applicable for describing a large class of coupled chaotic systems.

AT wants to thank A.~Altland, I.~L.~Aleiner and F.~W.~J.~Hekking for useful discussions. This research was supported by the SFB 237 of the DFG.

\end{document}